\documentclass[aps,prl,twocolumn,showpacs,groupedaddress]{revtex4}

\usepackage{amsmath}
\usepackage{graphicx}
\usepackage{dcolumn}
\usepackage{bm}

\begin{document}


\title{
Quantized vortices in $^{\bf 4}$He droplets: a quantum Monte Carlo study}

\author{E. Sola}
\author{J. Casulleras}
\author{J. Boronat}
\affiliation{Departament de F\'\i sica i Enginyeria Nuclear, Campus Nord
B4-B5, Universitat Polit\`ecnica de Catalunya, E-08034 Barcelona, Spain}

\date{\today}
\begin{abstract}
We present a diffusion Monte Carlo study of a vortex line excitation
attached to the center of a $^4$He droplet at zero temperature. The vortex
energy is estimated for droplets of increasing number of atoms, from $N=70$ up
to 300 showing a monotonous increase with $N$. The evolution of the core radius 
and its associated energy, the core energy, is also studied as a function of $N$. The core
radius is $\sim 1$ \AA\ in the center and increases when approaching the
droplet surface; the core energy per unit volume stabilizes at a value
2.8 K$\sigma^{-3}$ ($\sigma=2.556$ \AA) for $N \ge 200$.
\end{abstract}


\maketitle


Quantized vortices are one of the most outstanding signatures of
superfluidity. They have been widely observed in liquid $^4$He at
temperatures below the critical temperature $T_\lambda=2.17$ K~\cite{donnelly}. More recently,
vortices  have also been detected in dilute Bose~\cite{matthews} and
Fermi~\cite{zwierlein} gases when they
are magnetically trapped, their observation being considered the most clear 
indication of the achievement of their superfluid phases. $^4$He droplets    
produced in  free jet expansion experiments~\cite{Close98} are expected to
be also superfluid due to their very low temperature ($T=0.38$ K). 
However, in
this finite system the search for a direct signature of their superfluid
character is more difficult. On the one hand, indirect evidence of superfluidity has been
driven from the determination of the rotation spectrum of molecules
adsorbed into them~\cite{grebenev}. This very interesting phenomenon has been considered
the microscopic version of the famous Andronikashvili experiment. On the
other hand, the
detection of vortices in droplets would be an even more conclusive
proof of their superfluidity. Nevertheless, no signal 
of straight or circular vortices has been yet observed  in
experiments with $^4$He droplets, in spite of some theoretical arguments in favor 
of their possible metastability.

The stability of vortex excitations in $^4$He droplets was first studied by
Bauer \textit{et al.}~\cite{bauer}. They concluded that vortices are not
stable in droplets due to the high excitation energy required, compared to
the usual temperature at which they are produced. Later on, Lehmann and
Schmied~\cite{lehmann}, studying cold droplets, smaller 
than the ones analyzed in Ref.~\cite{bauer} were led to a different 
conclusion. They
stated that in small droplets, where only surface excitations (ripplons) are
relevant, vortices should be stable against decay. Finally,
density functional (DF) calculations~\cite{Dalfovo00,Ancilotto03} have predicted
that below a critical atomic number a linear vortex pinned to a dopant atom
or molecule can become stable, with a lifetime long enough to allow for  its
experimental detection. 

The excitation energy of a vortex line in a $^4$He droplet as a function of
its number of atoms $N$ is one of its more fundamental properties. An
accurate calculation of its value is crucial to elucidate its possible
stability and formation probability in jet expansions. To our
knowledge,
there is only one previous microscopic calculation of the excitation energy
associated to a vortex in a droplet. This calculation, performed with the path
integral Monte Carlo (PIMC) method~\cite{Draeger01}, was carried out for a $N=500$
$^4$He droplet and the energy obtained was more than a factor two smaller than
DF predictions~\cite{Dalfovo00,Ancilotto03}. In the present work, we present zero-temperature 
diffusion Monte Carlo (DMC) results of vortex energy and vortex structure 
in $^4$He droplets for different number of atoms.

A vortex is an excited state
of the $N$-particle Hamiltonian which corresponds to an eigenstate of the
angular momentum operator. Actually, it is an eigenstate of the $z$
component of the angular momentum, $L_z$, with eigenvalue $\hbar N l$  where
$l=1,2,\ldots$ is the quantum of circulation. In a general form, the
imaginary-time dependent wave function 
of a vortex in a quantum liquid can be written as
\begin{equation}
\Psi_v ({\bf{R}},t)=e^{i\Omega({\bf{R}},t)} \Phi({\bf{R}},t) \ .
\label{omega}
\end{equation}
By imposing that $\Psi_v$ is an eigenstate of the angular momentum
$L_z$,  Feynman \cite{Feynman55} obtained his famous
proposal  $\Omega({\bf{R}},t)=l \phi_{\rm F}({\bf{R}})$, with 
$\phi_{\rm F}({\bf{R}})=\sum_{j=1,N}\theta_j$, $\theta_j$ being the $j$-th polar
coordinate (in cylindric coordinates).

With the decomposition (\ref{omega}), the Schr\"odinger equation for
$\Psi_v ({\bf{R}},t)$ splits in two coupled equations, one for the modulus
\begin{equation}
-\frac{\partial\Phi}{\partial t} = D (\nabla \Omega)^2 \Phi - 
D\nabla  ^2 \Phi+ (V(\bf{R})-E) \Phi \ ,
\label{sch_1}
\end{equation}
and one for the phase
\begin{equation}
 \frac{\partial \Omega}{\partial t} = D \left[ \nabla ^2 \Omega + 
2(\nabla  \Omega)\frac{\nabla  \Phi}{\Phi} \right] \ ,
\end{equation}
with $D = \hbar^2/2m$. If a fixed form for the phase $\Omega$ is assumed, 
the equation for the 
modulus, Eq. (\ref{sch_1}), becomes the usual Schr\"odinger equation 
with one additional term (the first one on the right hand side). In this approximation,
known as fixed phase (FP), the vortex acts like a static external
potential. Using Feynman's expression for the phase, 
$\Omega=l \sum_{i=1}^N \theta_i$, this potential results
\begin{equation}
V_v({\bf{R}})=\sum_{i=1}^N\frac{\hbar^2 l^2}{2m \rho_i^2} \ ,
\label{Vvortex0}
\end{equation}
with $\rho_i$ the polar coordinate of particle $i$  in cylindric coordinates.
Therefore, in the FP approximation the problem of having a vortex inside 
the droplet is reduced to a different Hamiltonian ($ \tilde{H} = H + V_v$) in 
the Schr\"odinger equation to be stochastically solved. This approach was used in 
the PIMC calculation  of the vortex energy in a $^4$He droplet~\cite{Draeger01} and also 
recently in a Monte Carlo study of the excitation energy of vortices in trapped 
diluted Bose gases~\cite{mur-petit}.

In the present work, we will assume that the vortex line is fixed in the
$z$ direction of the center-of-mass (CM) reference system of the droplet. 
The vortex state is then an eigenstate of the $L_{z_{CM}}$ operator,
accounting for the translational invariance of the Hamiltonian.
In this case,  the  resulting 
potential ($\bar{V}_v$) that must be added to the
Hamiltonian in the droplet geometry is
\begin{equation}
\bar{V}_v({\bf{R}})=\frac{\hbar^2 l^2}{2m}\sum_{j=1}^N \left[ \frac{1}
{\rho_j^2}-\frac{1}{N}\sum_{k=1}^N \frac{\cos({\theta}_k-{\theta}_j)}
{\rho_k \rho_j}  \right] \ .
\label{VvortexCM}
\end{equation}
Coordinates in Eq. (\ref{VvortexCM}) and hereafter  
are referred to the center of mass. 
The potential $\bar{V}_v$ is very similar to the one for the bulk (\ref{Vvortex0}), 
but now written using CM coordinates, and a small correction of order 
$1/N$ is introduced. 

The FP approach to the excitation energy of a vortex line with 
the Feynman's phase could be thought as a too crude approximation for an
accurate microscopic treatment of the problem. An \textit{a priori} better
method would be to consider the superposition of  clockwise and anticlockwise
vortices, which are degenerate in energy, and using the resulting wave
function (orthogonal to the ground state) as a guiding wave function in
DMC. This leads to a Fermi-like problem due to the non-positivity of the
excited wave function which can be approximately solved in the fixed-node
(FN)
approximation. This method was used in Ref. \cite{Giorgini96} for studying
a vortex in a two-dimensional geometry. However, the results there obtained
showed that FN and FP predictions are almost compatible. Moreover, both Ref.
\cite{Giorgini96} and Ref. \cite{ortiz} studied the possible improvements upon
Feynman's phase by introducing backflow correlations on it and concluded
that their effect on the excitation energy is very small ($<1$\% in the
energy per particle). Therefore, it is sound to consider that
the FP method is also a good enough approximation to describe the vortex
attached to a droplet.

We have carried out DMC simulations of $^4$He droplets with number of atoms 
$N=70,128,200$, and $300$. To extract the excitation energy associated to
the vortex line we have calculated the energy of the droplets for both the
ground state and the droplet with a fixed vortex in the center of
mass. The ground-state trial wave function used for importance sampling in
the DMC method is a Jastrow form,
\begin{equation}
\Phi_0({\bf R}) =\prod_{i<j} \exp \left\{ {-\frac{1}{2}\left[ \left( \frac{b}
{r_{ij}} \right)^5 + \frac{\alpha^2 r_{ij}^2}{N} \right]} \right\} \ .
\label{phi0}
\end{equation}
The first term in the square bracket of Eq. (\ref{phi0}) is a McMillan
correlation factor accounting for  dynamical correlations induced by the
interatomic potential $V(r_{ij})$, and the second one is a gaussian
correlation to take into account the finite size of the droplet.

When the vortex is present, the Hamiltonian changes due to the potential
induced by the FP approximation (\ref{VvortexCM}). Therefore, we introduce
an additional correlation factor $f(\rho)$ to reduce the variance of the
energy estimation,
\begin{equation}
\Phi_v ({\bf R},t) = \Phi_0({\bf R}) \prod_{i=1}^N f(\rho_i) \ ,
\end{equation}
with $\rho_i$ the radial position of particle $i$. Among the different options
for $f(\rho)$, discussed for the bulk in Ref. \cite{Giorgini96}, we have
chosen 
\begin{equation}
f(\rho)= 1-e^{-\rho/a} 
\label{one-body}
\end{equation} 
with $a$ a variational parameter related to the \textit{radius} of the vortex core.
We have taken $a=1$ \AA, value which coincides with the old estimation of
the core radius by Rayfield and Reif~\cite{Rayfield64} and the more recent one
by Sadd \textit{et al.}~\cite{sadd}. 
This function satisfies $f(\rho)\rightarrow 0$  when $\rho \rightarrow 0$,
reflecting the repulsive character of the potential $\bar{V}_v$, and
approaches 1 far from the core. We have checked that the explicit form of
the function $f(\rho)$ satisfying both boundary conditions is not really
important and that, as expected, the energy of the system does not depend
on it.

The other two variational parameters $b$ and $\alpha$ appearing in the
trial wave function (\ref{phi0}) have been optimized by means of
variational Monte Carlo (VMC)
calculations. The optimal values are $b=1.19\sigma$ and
$\alpha=0.6\sigma^{-2}$  ($\sigma =2.556$ \AA) and their dependence with
the number of atoms $N$ is negligible in the range studied.

The main purpose of this work, i.e.,  the determination of the excitation energy
associated to a vortex line attached to the CM of the $^4$He droplet, 
 has been studied by other groups using mainly density
functional theory \cite{Ancilotto03,Dalfovo00,lehmann}.  Dalfovo \emph{et al.}
\cite{Dalfovo00}  proposed a liquid-drop formula for the energy
dependence on the number of particles,
\begin{equation}
E_v(N)=\lambda N^{1/3} + \beta N^{1/3} \log N + \gamma N^{-1/3} \ ,
\label{liquid_formula}
\end{equation}
$\lambda$, $\beta$ and $\gamma$ being parameters which were obtained by 
fitting Eq.
(\ref{liquid_formula}) to their DF results.  The final set of parameters was
$\lambda=2.868$ K, $\beta=1.445$ K and $\gamma =0.313 $ K.  

The specific
dependence on $N$ in Eq. (\ref{liquid_formula}) is derived using the 
hollow-core model. In this model,
the \emph{local} vortex energy is integrated over all the volume
occupied by particles
\begin{equation}
E_v(N) = \int_V dV \frac{1}{\rho^2} \ , 
\end{equation}
resulting in
\begin{equation}
E_v(N) = \frac{2\pi\hbar^2 D_0}{m} \left[ R \ln 
\left( \frac{2R}{a} \right) -R +  \frac{a^2}{2R } \right] \ .
\label{kv}
\end{equation}
Notice that one can rewrite Eq.  (\ref{kv}) in terms of $N$ just using
\mbox{$R=r_0 N^{1/3}$}, where \mbox{$r_0=\left(3/(4\pi
D_0)\right)^{1/3}$}, arriving in such a way to the same dependence on
$N$ as Eq. (\ref{liquid_formula}). However, the hollow-core method is a too simple
approximation to identify the parameters $\lambda$, $\beta$ and $\gamma$
using, for example, a constant density $D_0$.

\begin{table}[t]
\begin{center}
\begin{tabular}{l  c c c c}
\hline
 $ N             $&$ 70            $&$ 128        $&$ 200           $&$ 300          $ \\
\hline
$ E_0[K]         $&$   -223.6(2)   $&$ -497.4(5)  $&$ -861(1)    $&$ -1376(2)     $  \\
$ E_1[K]         $&$   -188.0(2)  $&$ -446.6(4) $&$  -799(1)   $&$ -1302(1)  $  \\
$ E_v[K]         $&$  35.7(3)   $&$ 50.8(6)    $&$   61(2)   $&$ 74(2)  $ \\
\hline
\end{tabular}
\caption{Total energies of $^4$He droplets with ($E_1$) and without ($E_0$) 
a vortex inside, and vortex excitation energy $E_v$. 
All energies are in units of Kelvin; in parenthesis, the statistical errors.}
\end{center}
\label{Vortextable}
\end{table}

Calculations performed using DF~\cite{Dalfovo00}  and MC 
evaluate the vortex energy as the difference between 
the total energies of the droplet with and without (ground-state) a vortex 
($ E_v(N) = E_1(N) - E_0(N) $).
In Table I, we present our DMC results of the energies $E_1(N)$, $E_0(N)$,
and $ E_v(N)$ for $^4$He droplets with $N=70, 128, 200$, and $300$.
The vortex energy $E_v$ comes from
the difference of two energies which increase with the number of atoms $N$
and therefore its statistical error also increases with $N$. 

\begin{figure}
\centerline{\includegraphics[width=0.7\linewidth,angle=-90]{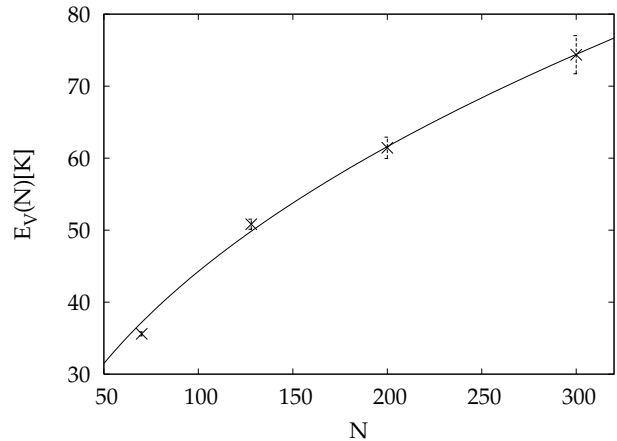}}
\caption{Vortex energies $E_v$ for $^4$He droplets with 
$N=70, 128, 200$ and $300$. 
The energy curve as a function of $N$ corresponds to the DF results from
Ref. \cite{Dalfovo00}. }
\label{vortexFig}
\end{figure}

The results contained in Table I show that the excitation energy $E_v$
increases monotonously with $N$. This can be more clearly observed in Fig. 
\ref{vortexFig} where the present  DMC results are compared with 
the liquid drop formula (\ref{liquid_formula}) as
reported by Dalfovo \emph{et al.} \cite{Dalfovo00}. One can observe that  
the DF estimation reproduces better our results for the larger droplets than for
the smaller ones.  This is what one a priori expects since  DF approximations
 work better for large droplets where application of mean field theory is more 
 justified; in spite of this, the difference is only $\sim$ 4\% for the
 smallest droplet
 studied ($N=70$).

\begin{figure}
\centerline{\includegraphics[width=0.7\linewidth,angle=-90]{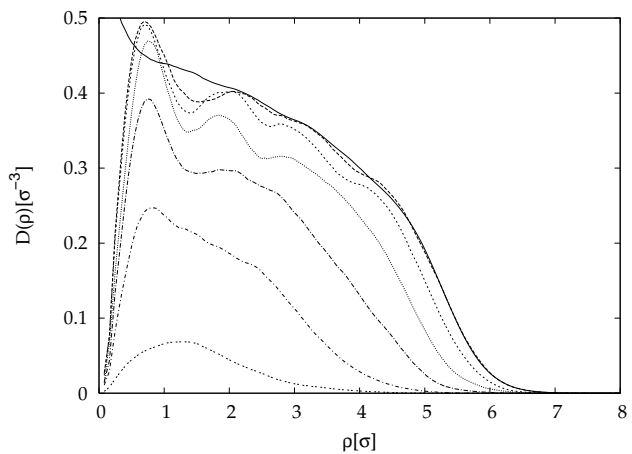}}
\caption{Density profiles for a $N=200$ $^4$He droplet with a vortex inside.
They are represented as a function of the radial coordinate $\rho$ and for
different slices corresponding to different $z$ values on the vortex axis.
From top to bottom the curves correspond to increasing values of $z$ from
$z=0$ in the center of the droplet up to $z$ values close to the surface. For
comparison, we also show the radial density profile of the droplet without the
vortex (solid line).
}
\label{vortexdensityFig}
\end{figure}

The repulsive potential induced by the vortex creates a \textit{hole} when
$\rho \rightarrow 0$ with a characteristic modulation of the density. This
behavior is shown in Fig. 2 where we have plotted the density profiles
$D(\rho)$ for different values of $z$ along the vortex axis. The slice
at $z=0$ corresponds to the one in the center of the droplet and is the
density profile with the highest peak. When $z$ increases, the radius of
the slice is
progressively smaller and also the oscillations are depressed as $z$
approaches the droplet radius. In the center of the vortex, the density goes to
zero in agreement with the DMC results derived en Ref. \cite{Giorgini96} for 
homogeneous 2D liquid $^4$He. A relevant parameter in the microscopic 
description of a vortex is the size of its core, what is called the core
radius $\xi$.   There is not a single definition of
$\xi$ but different definitions lead to quite similar results. We have used
the criterion of considering for $\xi$ the position of the maximum in the
azimuthal circulating current $J_\theta(\rho)$. If the vortex is described
by means of the Feynman approximation, as in the present work, it follows that
 $J_\theta(\rho)=D(\rho)/ \rho$. Therefore, we can get a direct estimation
 of $\xi$ from the density profiles $D(\rho)$ shown in Fig. 2. In the
 center of the droplet $\xi=1.0 \pm 0.1$ \AA\ and it increases  when $z$
 approaches the value of the droplet radius up to $\xi=1.4 \pm 0.1$ \AA. 

\begin{figure}
\centerline{\includegraphics[width=0.7\linewidth,angle=-90]{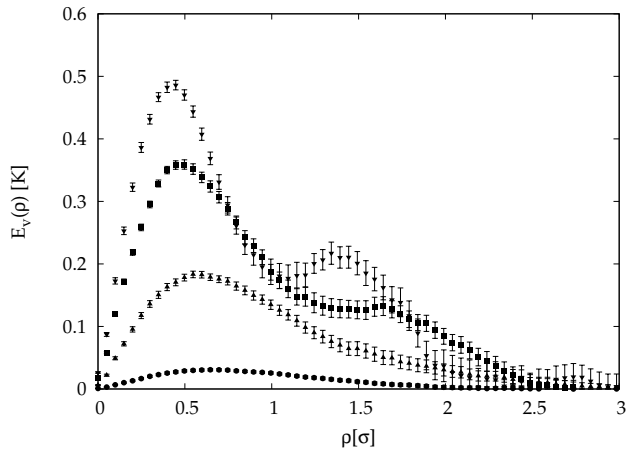}}
\caption{Local energy $E_v(\rho)$ for a $N=70$ $^4$He droplet with a vortex inside.
From top to bottom the curves correspond to increasing values of $z$ from
$z=0$ in the center of the droplet up to $z$ values close to the surface.
}
\label{vortexenergyFig}
\end{figure}


In a homogeneous system, and at large distances from the
vortex axis, the excitation vortex energy is usually decomposed in the form
\begin{equation}
E_v(\rho) = A \ln(\rho / \xi) + E_c
\label{hydrodyn}
\end{equation}   
with $A$ a constant, $E_c$  the core
energy, and $\rho$  the radial coordinate in cylindrical coordinates.
This model, which reproduces very well DMC results of the vortex energy in a 2D
geometry~\cite{Giorgini96}, splits the vortex energy in a constant term
$E_c$ associated to the \textit{hole} around the vortex axis and a
logarithmic term containing the hydrodynamic tail. In the case of droplets, the
behavior of the excitation energy is more complex due to their
inhomogeneity along the $z$ vortex axis and their finite size. 
In the function $E_v(\rho)$ one finds, superimposed to the monotonously
increasing law
(\ref{hydrodyn}), a decaying trend to zero when the surface of the
droplet is reached. This behavior is shown in Fig. 3 where we have plotted the
 function  $E_v(\rho)$ for a $N=70$ $^4$He droplet. The function $E_v(\rho)$
 shows clearly a peak corresponding to the core of the vortex, especially
 for the inner slice, and a decay to zero in the surface. The size of
 the droplets studied is too small to see the signature of the hydrodynamic
 tail but we can estimate the energy of the core $E_c$ by summing the
 local energy up to the estimated vortex core radius $\xi$. Extending this sum
 along all the vortex axis we obtain an energy which increases with $N$:
  $E_c=9.74(4)$ K and 19.2(3) K for $N=70$ and 300, respectively.
 Normalizing $E_c$ with respect to the \textit{volume} of the core, the core
 energy per volume unit approaches a constant value of 2.8 K$\sigma^{-3}$ for
 droplets with $N \ge 200$.

Summarizing, we have carried out the first microscopic calculation of the
properties of a vortex line attached to a $^4$He droplet at zero temperature. 
The energies
obtained are in good agreement with DF estimations and therefore give
additional confidence on their predictions. Moreover, the magnitude of the core
radius and core energy has also been studied for the first time in these
inhomogeneous system. The core radius is $\sim 1$ \AA\ in the center and increases 
when approaching the
droplet surface; the core energy per unit volume stabilizes at a value
2.8 K$\sigma^{-3}$   for $N \ge 200$.
We hope our results can help and stimulate further 
experimental work towards the long-standing purpose of detecting vortex
excitations in $^4$He droplets.

We thank Manuel Barranco and Marti Pi for useful discussions.
We acknowledge financial support from DGI (Spain) Grant No.
FIS2005-04181 and Generalitat de Catalunya Grant No. 2005SGR-00779.


\begin{thebibliography}{21}


\bibitem{donnelly} R. J. Donnelly, \textit{Quantized Vortices in Helium II}
(Cambridge University Press, Cambridge, 1991).

\bibitem{matthews} M. R. Matthews, B. P. Anderson, P. C. Haljan, D. S.
Hall, C. E. Wieman, and E. A. Cornell, Phys. Rev. Lett. \textbf{83}, 2498
(1999).

\bibitem{zwierlein} M. W. Zwierlein, J. R. Abo-Shaeer, A. Schirotzek, C. H.
Schunck, and W. Ketterle, Nature \textbf{435}, 1047 (2005).


\bibitem{Close98} J. D. Close, F. Federman, K. Hoffman, 
and N. Quaas,  J. Low Temp. Phys. {\bf 111}, 661 (1998).

\bibitem{grebenev} S. Grebenev, J. P. Toennies, and A. F. Vilesov, Science
\textbf{279}, 2083 (1998).


\bibitem{bauer} G. H. Bauer, R. J. Donnelly, and W. F. Vinen, J. Low Temp.
Phys. \textbf{98}, 47 (1995). 

\bibitem{lehmann} K. K. Lehmann and R. Schmied, Phys. Rev. A \textbf{68},
224520 (2003).

\bibitem{Dalfovo00} F. Dalfovo, R. Mayol, M. Pi, and M. Barranco, 
Phys. Rev. Lett. {\bf 85}, 1028 (2000). 

\bibitem{Ancilotto03} F. Ancilotto, and M. Barranco and M. Pi, 
Phys. Rev. Lett. {\bf 91}, 105302 (2003).  
 

\bibitem{Draeger01} E. W. Draeger, Ph.D. thesis, University of Illinois at
Urbana-Champaign (2001). 

\bibitem{Feynman55} R. P. Feynman, in  {\it Progress in Low Temperature Physics I}, 
Ed. C.J.Goster (North-Holland, Amsterdam, 1955) Chap. II. 


\bibitem{mur-petit} J. K. Nilsen, J. Mur-Petit, M. Guilleumas, M.
Hjorth-Jensen, and A. Polls, Phys. Rev. A \textbf{71}, 053610 (2005).


\bibitem{Giorgini96} S. Giorgini, J. Boronat, and J. Casulleras, 
Phys. Rev. Lett. {\bf 77}, 2754 (1996).

\bibitem{ortiz} G. Ortiz and D. M. Ceperley, Phys. Rev. Lett. \textbf{75},
4642 (1995).

\bibitem{Rayfield64}  G. W. Rayfield and F. Reif, Phys. Rev. 
{\bf 136}, A1194 (1964). 

\bibitem{sadd} M. Sadd, G. V. Chester, and L. Reatto, Phys. Rev. Lett.
\textbf{79}, 2490 (1997).

\end{thebibliography}
\end{document}